\documentclass[twocolumn]{aastex631}
\usepackage[utf8]{inputenc}
\usepackage{amsmath,amsxtra,amssymb,latexsym,amscd,amsthm}
\usepackage{mathrsfs}
\usepackage[mathscr]{eucal}
\usepackage{mathrsfs}
\usepackage{makeidx}
\usepackage{soul}
\usepackage{graphicx}
\usepackage{subfigure}
\usepackage{booktabs}
\usepackage{multirow}
\usepackage{threeparttable}
\usepackage{hyperref}

\shorttitle{VSI \& SI \& RWI \& KHI}
\shortauthors{Huang \& Bai}
\graphicspath{{./}{figures/}}

\begin{document}

\title{Dust Clumping in Outer Protoplanetary Disks: the Interplay Among Four Instabilities}

\author[0000-0002-7575-3176]{Pinghui Huang \dag}
\affiliation{CAS Key Laboratory of Planetary Sciences, Purple Mountain Observatory, Chinese Academy of Sciences, Nanjing 210008, People’s Republic of China}
\affiliation{Department of Physics \& Astronomy, University of Victoria, Victoria, British Columbia, V8P 5C2, Canada}
\affiliation{Institute for Advanced Study, Tsinghua University, Beijing 100084, People’s Republic of China}

\author[0000-0001-6906-9549]{Xue-Ning Bai \ddag}
\affiliation{Institute for Advanced Study, Tsinghua University, Beijing 100084, People’s Republic of China}
\affiliation{Department of Astronomy, Tsinghua University, Beijing 100084, People’s Republic of China}

\correspondingauthor{Pinghui Huang; Xue-Ning Bai}
\email{phhuang@pmo.ac.cn; xbai@tsinghua.edu.cn}

\begin{abstract}
Dust concentration in protoplanetary disks (PPDs) is the first step towards planetesimal formation, a crucial yet highly uncertain stage in planet formation. Although the streaming instability (SI) is widely recognized as a powerful mechanism for planetesimal formation, its properties can be sensitive to the gas dynamical environment. The outer region of PPDs is subject to the vertical shear instability (VSI), which could further induce the Rossby wave instability (RWI) to generate numerous vortices. In this work, we use the multifluid dust module in Athena++ to perform a 3D global simulation with mesh refinement to achieve adequate domain size and resolution to resolve and accommodate all these instabilities. The VSI mainly governs the overall gas dynamics, dominated by the breathing mode due to dust mass loading. The dust strongly settles to the midplane layer, which is much more densely populated with small vortices compared to the dust-free case. Strong dust clumping is observed, which is likely owing to the joint action of the SI and dusty RWI, and those sufficient for planetesimal formation reside only in a small fraction of such vortices. Dust clumping becomes stronger with increasing resolution, and has not yet achieved numerical convergence in our exploration. In addition, we find evidence of the Kelvin-Helmholtz instability (KHI) operating at certain parts of the dust-gas interface, which may contribute to the temporary destruction of dust clumps.
\end{abstract}

\section{Introduction}~\label{sec:intro}

Planetesimal formation marks the initial stage of planet formation, yet it remains one of the least understood phases~\citep{ChiangYoudin2010}. This is not only due to the difficulty in directly observing the process, but also because of significant uncertainties in the theoretical models. It is widely accepted that planetesimal formation proceeds in a two-stage process: first, dust particles concentrate into dense clumps, and then these clumps undergo gravitational collapse due to their own self-gravity. The first stage is the most critical, where it is generally understood as a result of the Streaming Instability~\citep[SI,][]{GoodmanPindor2000,YoudinGoodman2005} or dust trapping in pressure maxima~\citep{Weidenschilling1980Planetesimals,BargeSommeria1995}. However, both scenarios sensitively depend on the gas dynamics in the protoplanetary disk (PPD).

PPDs are believed to be (weakly) turbulent. This turbulence not only influences the overall structure of the disks but also introduces diffusive effects that strongly affect dust dynamics. In the vertical direction, the turbulence properties determines the vertical profile of dust particles~\citep{Cuzzi1993,YoudinLithwick2007}, while in the horizontal direction, many gas dynamical processes can lead to the formation of pressure bumps or anti-cyclonic vortices that can efficiently trap dust~\citep{Bae2023}. Moreover, the SI is mostly studied under idealized conditions without external turbulence. There has been controversy on role of external turbulence. On one hand, turbulence is expected to suppress the SI based on linear theory~\citep{Umurhan2020,ChenLin2020} and simulation of driven (forced) turbulence~\citep{Gole2020,Lim2024}. However, the situation can be very different under more realistic gas dynamics.

Here we focus on the outer region of PPDs (at a distance of $\gtrsim10$ AU). On one hand, PPDs are weakly ionized, leading to weak coupling between gas and magnetic field, which is mainly governed by ambipolar diffusion (AD) in outer disk conditions. The magnetorotational instability~\citep[MRI;][]{BalbusHawley1998}, a powerful mechanism for driving turbulence in most accretion disks, is expected to be damped by AD, leading to weak-to-modest turbulence~\citep{BaiStone2011,Simon2013,CuiBai2021}. On the other hand, the vertical shear instability~\citep[VSI,][]{Nelson2013,BarkerLatter2015} is expected to operate thanks to the relatively short cooling timescale in outer PPDs, generating moderate level of turbulence. We note that in reality, the MRI and VSI can co-exist~\citep{CuiBai2022}, with turbulence exhibiting characteristic properties of both. In this work, we focus on the VSI turbulence and its interplay with dust dynamics.

Recently in~\cite{HuangBai2025I} (here after HB25), we investigated the dust dynamics in both 2D and 3D VSI turbulence in a suite of global simulations, including dust back-reaction. Our 2D simulations demonstrated that SI can coexist with VSI in axisymmetric setups, where weak pressure bumps generated by 2D VSI turbulence may help form additional dust clumps, consistent with the findings of~\cite{SchaferJohansen2020}, ~\cite{SchaferJohansen2022} and~\cite{Schafer2025}. In 3D, the VSI turbulence can produce zonal flows that trigger the Rossby wave instability~\cite[RWI,][]{LovelaceLi1999,LiFinn2000,LiColgate2001}, leading to the formation of anticyclonic vortices. With dust, such vortices can substantially enhance dust trapping~\citep{Richard2016,MangerKlahr2018,Manger2020}. Our study also showed that dust feedback significantly affects 3D VSI turbulence, where dust buoyancy/mass loading suppresses VSI corrugation modes, allowing dust to settle deeper, creating more favorable conditions for the SI to trigger planetesimal formation. However, the resolution in our earlier 3D simulations were insufficient to resolve the SI and capture dust clumping. We expect that with higher resolution, the SI and/or dusty RWI~\citep{LiuBai2023} should operate in the dust layer under 3D VSI turbulence.

In this study, we extend the 3D global simulation of HB25 by incorporating several additional levels of mesh refinement. This allows us to, for the first time, simultaneously resolve the SI while accommodating the VSI turbulence in a 3D global setup. We expect that this setup will enable us to study the interplay of the VSI and RWI with the SI towards a better understanding of how dust clumping and planetesimal formation occur under more realistic gas dynamics in the outer PPDs. Moreover, a thin dust layer can be further subject to the Kelvin-Helmholtz instability (KHI), as a result of strong vertical shear in azimuthal velocity at the interface of the dust layer~\citep{Sekiya1998,Johansen2006}. This is expected to generate additional stirring to sustain the dust layer thickness. As a result, our simulation allows us to simultaneously observe the VSI, SI, RWI and KHI.

The outline of this paper is as follows. We describe the numerical setup in Section~\ref{sec:method}. In Section~\ref{sec:simulations}, we present the results of 3D global simulation with mesh refinement. In Section~\ref{sec:summary}, we provide the conclusions and the further discussions. In HB25, we have already investigated the interplay between VSI and SI, as well as VSI and RWI. For readers who may not be familiar with these topics, we provide a brief introduction about VSI, SI and RWI in Appendix~\ref{app:background}.

\begin{figure*}[htp]
\centering
\includegraphics[scale=1.0]{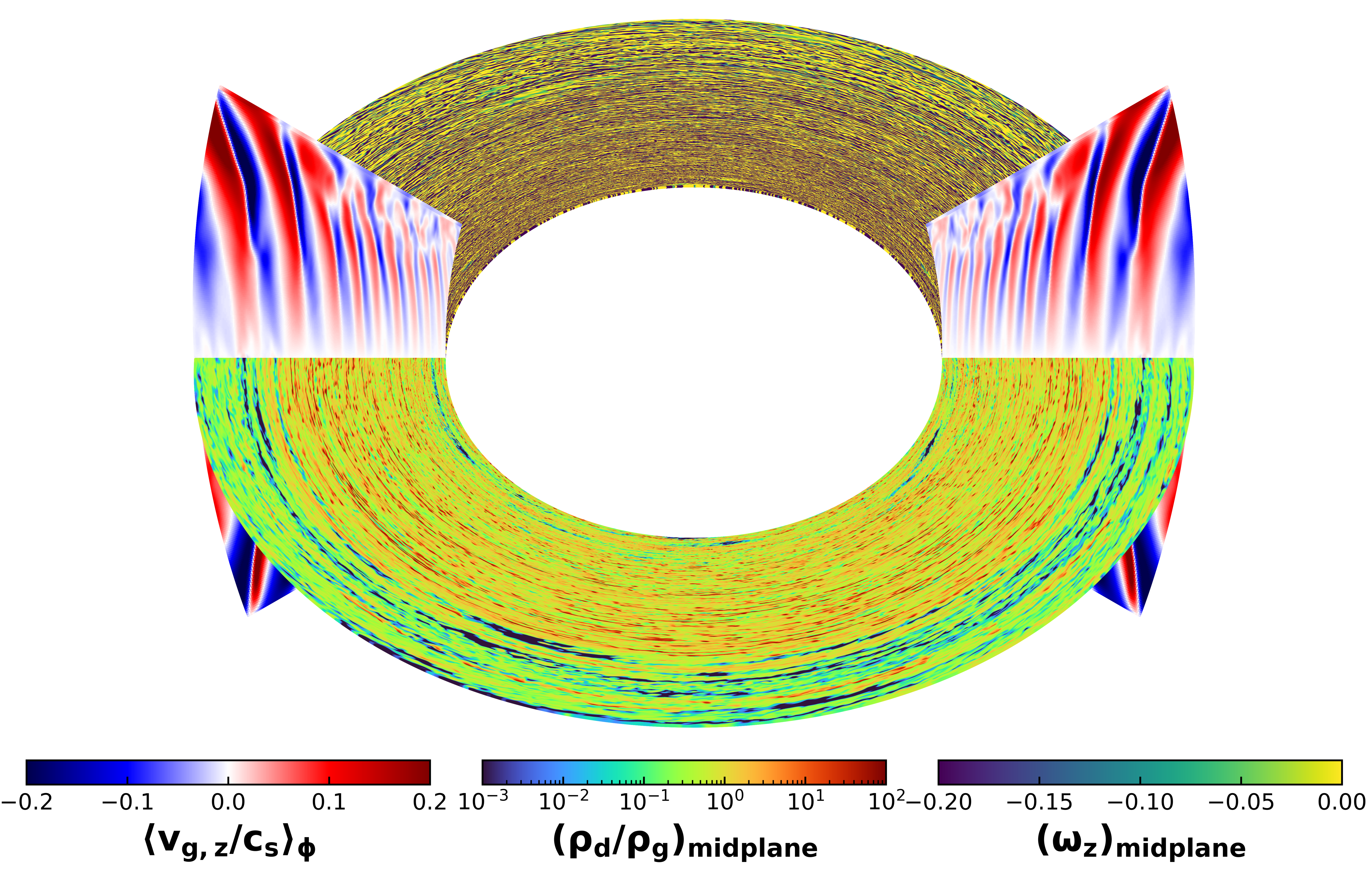}
\caption{A 3D schematic diagram of the snapshot at 550 orbits for the ``3D-FIs'' model. The two vertical slices represent the azimuthally averaged vertical Mach number $\langle v_{\text{g},z}/c_\text{s} \rangle_\phi$, the lower and upper semi-circles show the dust-gas density ratio at the midplane $(\rho_\text{d}/\rho_\text{g})_\text{midplane}$, and the vertical vorticity at the midplane $(\omega_z)_\text{midplane}$, respectively. For visualization purposes, the radial domain is limited to $r \in [1.2, 2.4]$.}
\label{fig:3D_general}
\end{figure*}

\begin{figure*}[htp]
\centering
\includegraphics[scale=0.55]{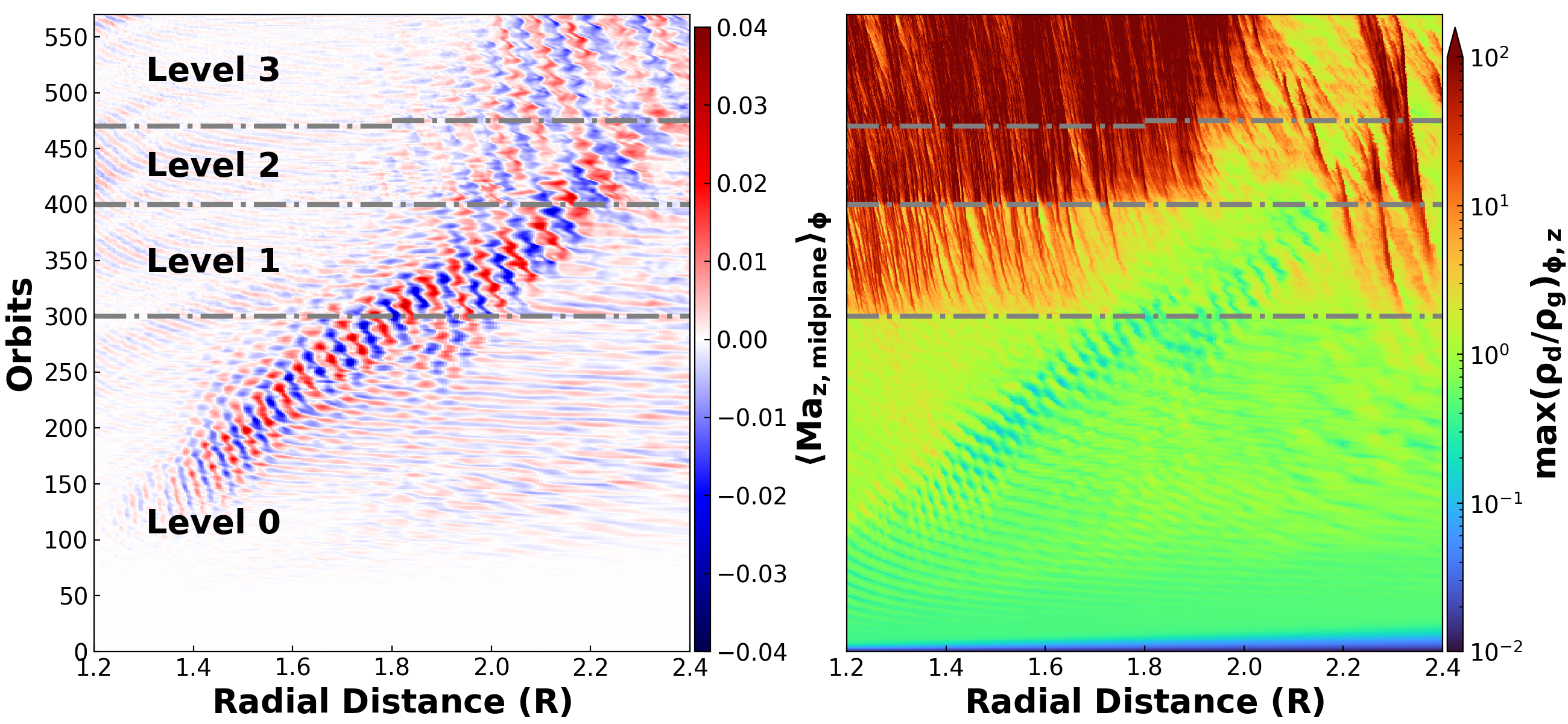}
\caption{The space-time ($R-t$) plots of the azimuthally averaged vertical Mach number at the midplane (left: $\langle Ma_{z,\text{midplane}} \rangle_\phi \equiv \langle v_{\text{g},z,\text{midplane}}/c_\text{s} \rangle_\phi$); and the maxima of the dust-gas density ratio (right: $\max(\rho_\text{d}/\rho_\text{g})_{\phi,z}$) for the ``3D-FIs'' model are shown. The grey dashed-dotted lines divide the stages where the maximum level in the dust layer is achieved. Level 0 spans 0 to 300 orbits, level 1 spans 300 to 400 orbits, level 2 spans 400 to 470 orbits, and level 3 spans 470 to 570 orbits.}
\label{fig:3D_space_time_3D_AMR}
\end{figure*}

\section{Numerical Setups}~\label{sec:method}

We use the multifluid dust module~\citep{HuangBai2022} in the Athena++ code~\citep{Stone2020} to conduct a 3D global simulation in spherical-polar coordinates ($r-\theta-\phi$) with mesh refinement for both gas and dust. This simulation explores the interaction between VSI and SI, with RWI and KHI emerging as by-products. We name this simulation ``3D-FIs'' (Four Instabilities) to reflect the interplay among these instabilities.

The computational domain spans $r \in [1,3]$, $\theta \in [\pi/2 - 0.4, \pi/2 + 0.4]$, and $\phi \in [0, \pi]$. The setup follows the 2D and 3D simulations described in HB25, using a locally isothermal equation of state, $P = c_\text{s}^2(R) \rho_\text{g}$, where $P$ and $c_\text{s}$ are gas pressure and sound speed, respectively. The radial profiles of sound speed, as well as initial gas and dust densities ($\rho_\text{g,init},\rho_\text{d,init}$), are given by
\begin{equation}
c_\text{s}^2(R) = c_\text{s,0}^2 \left(\frac{R}{R_0}\right)^{-\frac{1}{2}},
\end{equation}
\begin{equation}
  \rho_\text{g,init}(R,z) = \rho_0 \left(\frac{R}{R_0}\right)^{-\frac{3}{2}} \exp \left[\frac{GM}{c_\text{s}^2}\left(\frac{1}{r}-\frac{1}{R}\right) \right],\\
\end{equation}
\begin{equation}
  \rho_\text{d,init} = 0.01 \rho_\text{g,init}.
\end{equation}
where $GM \equiv 1 $ for stellar gravity, and we take $ \rho_0 \equiv 1 $, $c_{\text{s},0} = 0.08$ and $R_0 \equiv 1$ in this simulation. A single dust species with a constant Stokes number $St \equiv T_\text{s}\Omega_\text{K} = 0.1$ is considered, where $T_\text{s}$ is the dust stopping time. Governing equations and initial/boundary conditions are detailed in Sections 2.1 and 2.2 of HB25. We use four stages of refinement to achieve three refinement levels, optimizing memory use and load balancing across computational nodes. The activation timing of mesh refinement does not significantly affect the final results.

We adopt the same ``3D-FB'' setup following Table 1 of HB25 as the root-grid (level 0) configuration for the ``3D-FIs'' model in this study. The root-grid resolution for both runs is $512 \times 256 \times 768$ cells in the $r$--$\theta$--$\phi$ directions. This corresponds to an effective resolution of approximately $40 \times 28 \times 22 \simeq 29^3$ cells per $H_\text{g}^3$ at $r = 1.5$. There is no initial difference in the numerical setup between the ``3D-FB'' and ``3D-FIs'' models. Dust is initially released, and dust feedback is included in both cases. However, in ``3D-FB'' of HB25, mesh refinement was not enabled, and thus the SI in ``3D-FB'' could not be resolved. From 300 to 400 orbits of ``3D-FIs'', we activate the first level of mesh refinement (doubling the resolution) to resolve the thin dust layer near the midplane. Subsequent refinement transitions proceed sequentially: from level 1 to level 2, and then from level 2 to level 3. During the transition from level 2 to level 3, mesh refinement first focuses on the region $r = 1.2 -1.8$, followed by refinement in the region $r = 1.8 - 2.4$. After full three-level mesh refinement in ``3D-FIs'', the effective resolution reaches $318 \times 227 \times 173 \simeq 232^3$ cells per $H_\text{g}^3$ at $r = 1.5$ and $\theta = \pi/2$ (midplane), which is sufficient to capture the unstable modes of SI.

The finest grids of ``3D-FIs'' cover $r \in [1.2, 2.4]$, $\theta \in [\pi/2 - 0.025, \pi/2 + 0.025]$, and $\phi \in [0, \pi]$. Each snapshot with full three-level refinement involves 1.2 billion grid points. Due to the computational cost of three refinement stages for ``3D-FIs'' ($\sim 1.5 \times 10^4$ CPU*hours per orbit), we limit it to 100 orbits between 470 and 570 orbits.

For diagnostics, we use cylindrical coordinates ($R-\phi-z$) to analyze and present selected results of ``3D-FIs'', following the methods in Section 2.3 and Appendix A of HB25. Consistent with HB25, final outcomes remain unaffected by whether dust is introduced initially or after VSI turbulence saturation.

\section{Numerical Results, the Coexistence of VSI, SI, RWI and KHI}~\label{sec:simulations}

In this section, we present the main simulation results. In particular, we closely compare the results of our new ``3D-FIs'' model with the ``3D-FB'' model in HB25 following similar analysis procedures and diagnostics, highlighting the new physics brought by the higher-resolution thanks to mesh refinement.

\begin{figure*}[htp]
\centering
\includegraphics[scale=0.60]{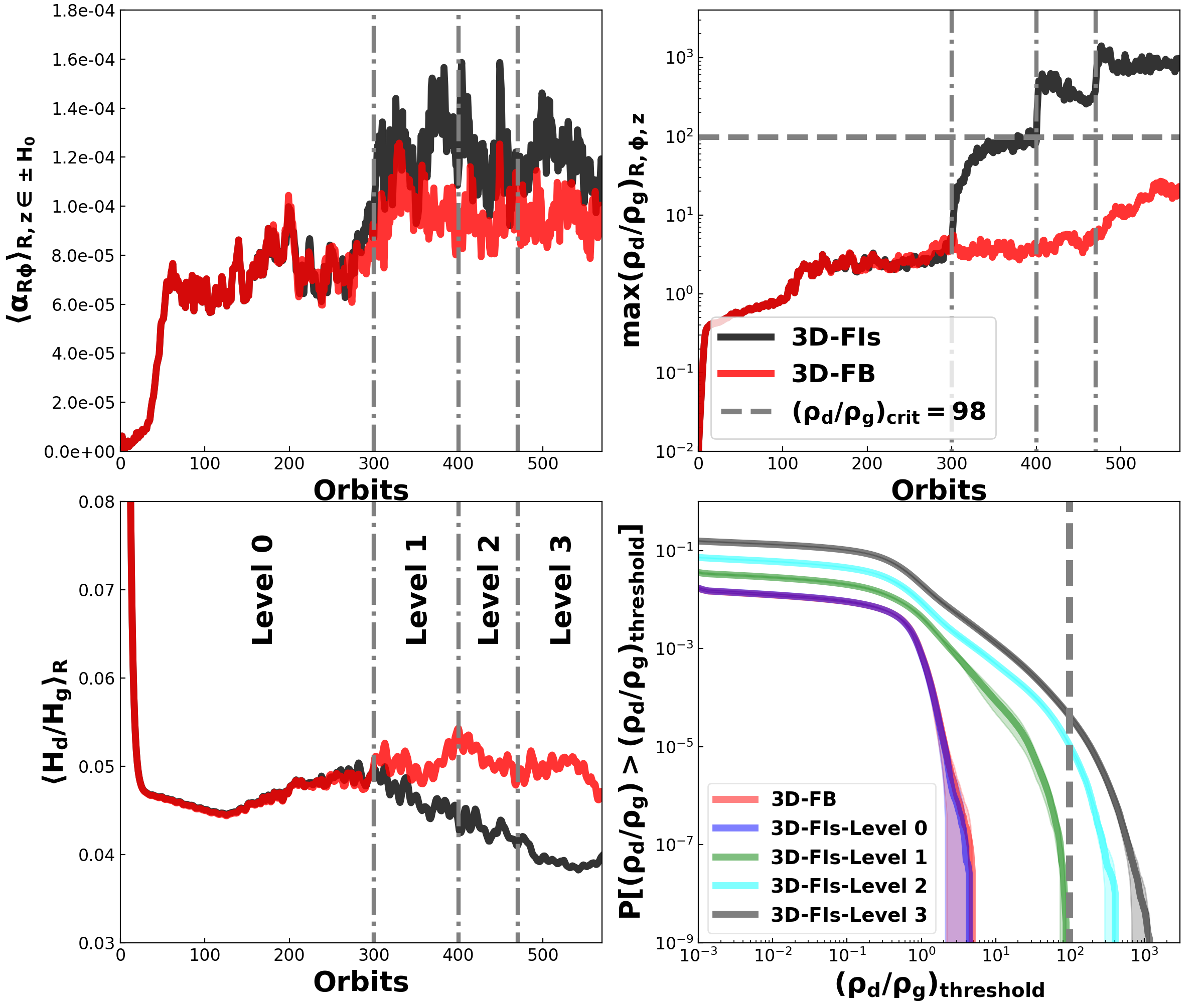}
\caption{Top-left: Temporal evolution of the radially and vertically averaged Reynolds stress, $\langle \alpha_{R\phi} \rangle_{R,z \in \pm H_0}$, where $H_0 = 0.08$. Top-right: Temporal evolution of the maximum dust-gas density ratio, $\max(\rho_\text{d}/\rho_\text{g})_{R,\phi,z}$. Bottom-left: Temporal evolution of the radially averaged dust-gas scale height ratio, $\langle H_\text{d}/H_\text{g} \rangle_R$. Bottom-right: Cumulative distribution function (CDF) of dust-gas density ratios, computed by counting the number of grid cells, $P[(\rho_\text{d}/\rho_\text{g}) > (\rho_\text{d}/\rho_\text{g})_\text{threshold}]$. The red solid lines represent the properties of the ``3D-FB'' model in HB25 without mesh refinement (measured at the same period as we measure the level 0 CDF), while the black solid lines correspond to the ``3D-FIs'' model. The gray dashed lines indicate the (rough) critical dust-gas density ratios required for planetesimal formation. The blue, green, cyan and black lines in the CDF plot indicate different refinement stages of ``3D-FIs'' model. Radially averaged values are calculated within the range $R \in [1.2, 2.4]$. Similar to Figure~\ref{fig:3D_space_time_3D_AMR}, the gray dashed-dotted lines indicate the stages with the maximum refinement level.}
\label{fig:3D_temperoal_evolution}
\end{figure*}

\subsection{Overview of Simulation Results}~\label{subsubsec:general}

We start from Figure~\ref{fig:3D_general} that illustrates the overall features of the ``3D-FIs'' simulation from the snapshot at 550 orbits. The VSI generates strongly anisotropic turbulence throughout the entire simulation domain, characterized by strong vertical motion dominated by the breathing modes (HB25). This leads to stronger dust settling than in 2D, and enhanced dust-gas density ratio at the midplane region. In the meantime, the dusty RWI stems from the zonal flows produced by the VSI, produces numerous small vortices, many of which are dust clumps that contain high dust-gas density ratios ($\rho_\text{d}/\rho_\text{g} \gtrsim 100$) at the midplane (see Section~\ref{subsubsec:dust}).

The overall properties of the VSI turbulence is largely unaffected by the refinement of the dust layer. This can be seen from the left panel of Figure~\ref{fig:3D_space_time_3D_AMR}, which shows the space-time ($R-t$) plot of the vertical Mach number at the midplane, $\langle Ma_{z,\text{midplane}} \rangle_\phi$. It closely resembles that of the ``3D-FB'' model in HB25, which shows ``trains'' of inward-propagating inertial waves, while the transition between two neighboring wave zones propagates outward at the group velocity of the inertial waves. No abrupt change is observed to the wave patterns as additional refinement level is introduced.

On the other hand, dust clumping is highly sensitive to resolution, as shown in the right panel of Figure~\ref{fig:3D_space_time_3D_AMR} that illustrates the maxima of the dust-gas density ratios, $\max{(\rho_\text{d}/\rho_\text{g})_{\phi,z}}$. The clumping phenomena occurs right after turning on the first level of mesh refinement, and every further refinement leads to a leap in $\max{(\rho_\text{d}/\rho_\text{g})_{\phi,z}}$ without sign of convergence (see further discussion in Section~\ref{subsubsec:dust}). All dust clumps slightly migrate inward. According to the linear eigenvalue analysis, the most unstable modes for SI with $St = 0.1$ exhibit growth rates of $0.1\text{--}0.4\;\Omega_\text{K}^{-1}$ at the finest resolution used in this study~\cite[see Figure 1 in ][]{YoudinJohansen2007}. When full three-level mesh refinement is enabled, the effective resolution reaches $\gtrsim 200^3$ cells per $H_\text{g}^3$, allowing these fastest-growing modes to be well resolved. As a result, strong dust concentration emerges rapidly, within just a few orbits—consistent with the predicted SI growth rates.

More quantitatively, Figure~\ref{fig:3D_temperoal_evolution} represents various parameters in the ``3D-FIs'' simulation, with those from the ``3D-FB'' model in HB25. These parameters include: the temporal evolution of the Reynolds stress, $\langle \alpha_{R\phi} \rangle_{R, z \in \pm H_0}$ (as defined in Equation~14 of HB25), the maxima of the dust-gas density ratios, $\max{(\rho_\text{d}/\rho_\text{g})_{R,\phi,z}}$, and the radially averaged dust-gas scale height ratio, $\langle H_\text{d}/H_\text{g} \rangle_R$ (Equation~22 of HB25). Their initial evolution are identical to those of the ``3D-FB'' model during the first 300 orbits. When mesh refinement is activated gradually after 300 orbits, stronger Reynolds stress emerges near the midplane, indicating that additional instabilities (e.g., SI, dusty RWI, and KHI) develop within the global VSI turbulence. The progressively higher dust-gas density ratios ($\max{(\rho_\text{d}/\rho_\text{g})} = 100 \sim 1000$) with increasing resolution suggests that the SI begins to be resolved and generates dust clumps. Although not yet reaching convergence, the dust-gas density ratios within these clumps can well exceed the critical value $(\rho_\text{d}/\rho_\text{g})_\text{crit}\sim100$ (Equation~24 in HB25) in typical outer disk conditions after reaching our second finest resolution, potentially allowing dust to gravitationally collapse into planetesimals with the aid of self-gravity. Given that the properties of the VSI turbulence largely remain similar in both ``3D-FB'' and ``3D-FIs'', this reduction in $H_\text{d}/H_\text{g}$ in ``3D-FIs'' is associated with dust clumping itself which mostly occurs in the midplane region.

\subsection{Dust Clumps and Vorticies}~\label{subsubsec:dust}

\begin{figure*}[htp]
\centering
\includegraphics[scale=0.40]{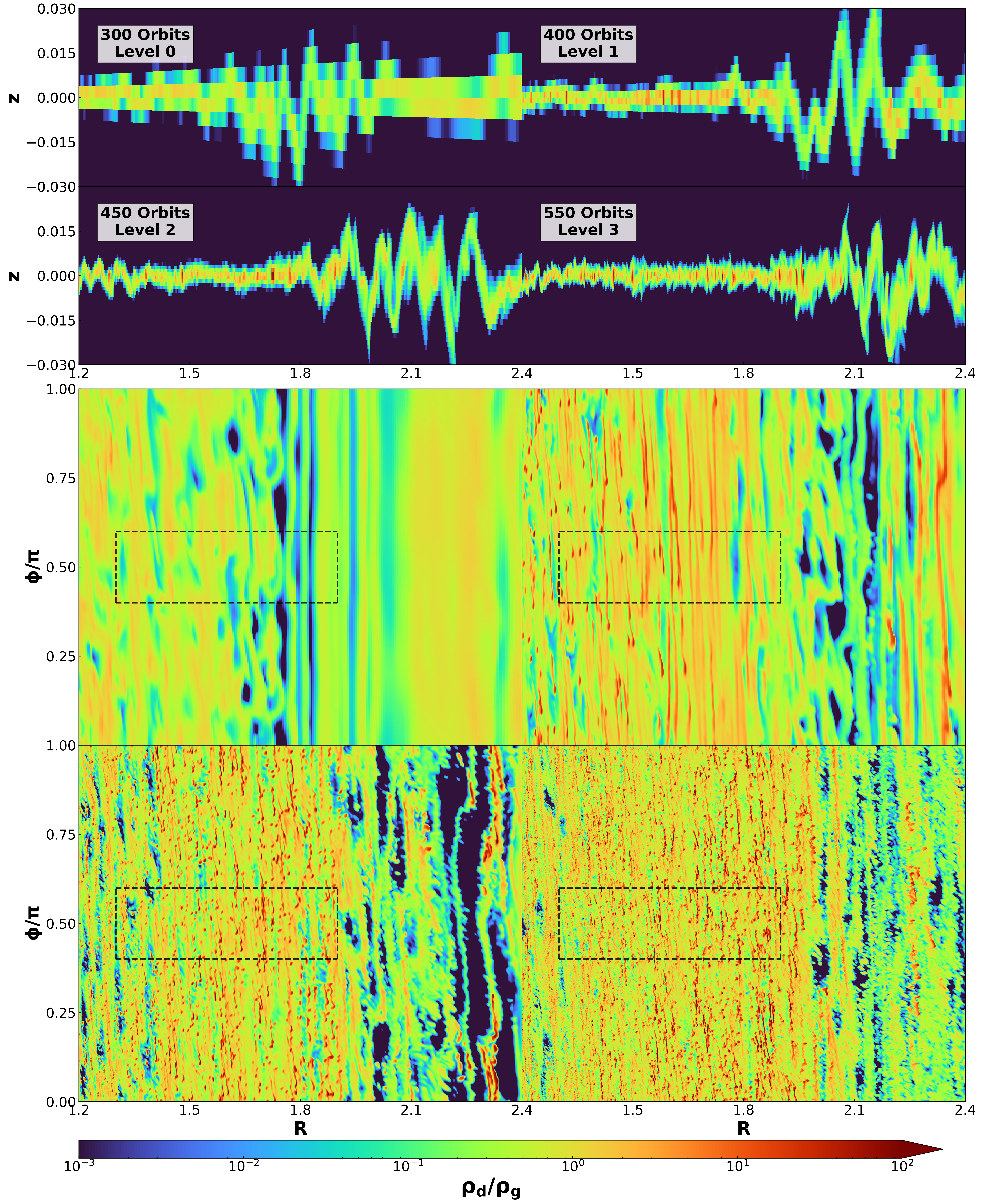}
\caption{The dust-gas density ratios $\rho_\text{d}/\rho_\text{g}$ on the $R$-$z$ ($\phi = \pi/2$) section (top four panels) and $R$-$\phi$ ($\theta = \pi/2$, $z = 0$, midplane) section (bottom four panels) for the ``3D-FIs'' model at various times are shown. From left to right, from top to bottom in the top four panels, the snapshots are taken at 300, 400, 450, and 550 orbits. The bottom four panels correspond to the same times as the top four panels. Different times and the maximum mesh levels are marked in the top-left corners of the top four panels. The black dashed rectangles indicate the zoomed-in regions shown in Figure~\ref{fig:3D_vorticity_ratio_3D_AMR}.}
\label{fig:3D_dust_ratio_3D_AMR}
\end{figure*}

\begin{figure*}[htp]
\centering
\includegraphics[scale=0.35]{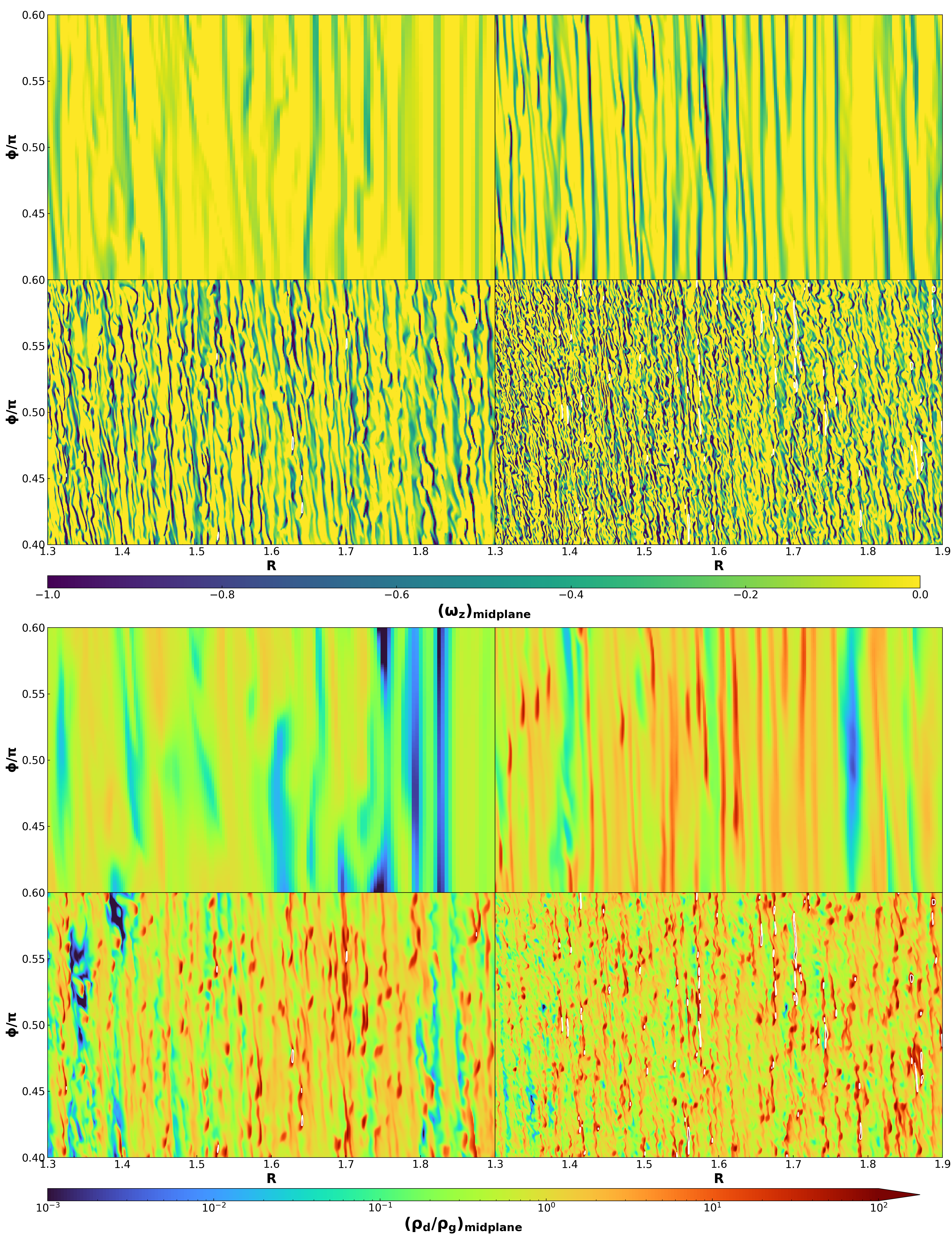}
\caption{The zoomed-in plots of gas vertical vorticity at the midplane $(\omega_z)_\text{midplane}$ (top four panels) and dust-gas density ratios at the midplane $(\rho_d/\rho_g)_\text{midplane}$ (bottom four panels) at various snapshots. The order of snapshots is same as Figure~\ref{fig:3D_dust_ratio_3D_AMR}. The white spots and contours indicate the regions where the local dust-gas density ratios exceed the critical ratio $(\rho_\text{d}/\rho_\text{g})_\text{crit} = 98$ (Equation 24 in HB25).}
\label{fig:3D_vorticity_ratio_3D_AMR}
\end{figure*}

\begin{figure*}[htp]
\centering
\includegraphics[scale=0.35]{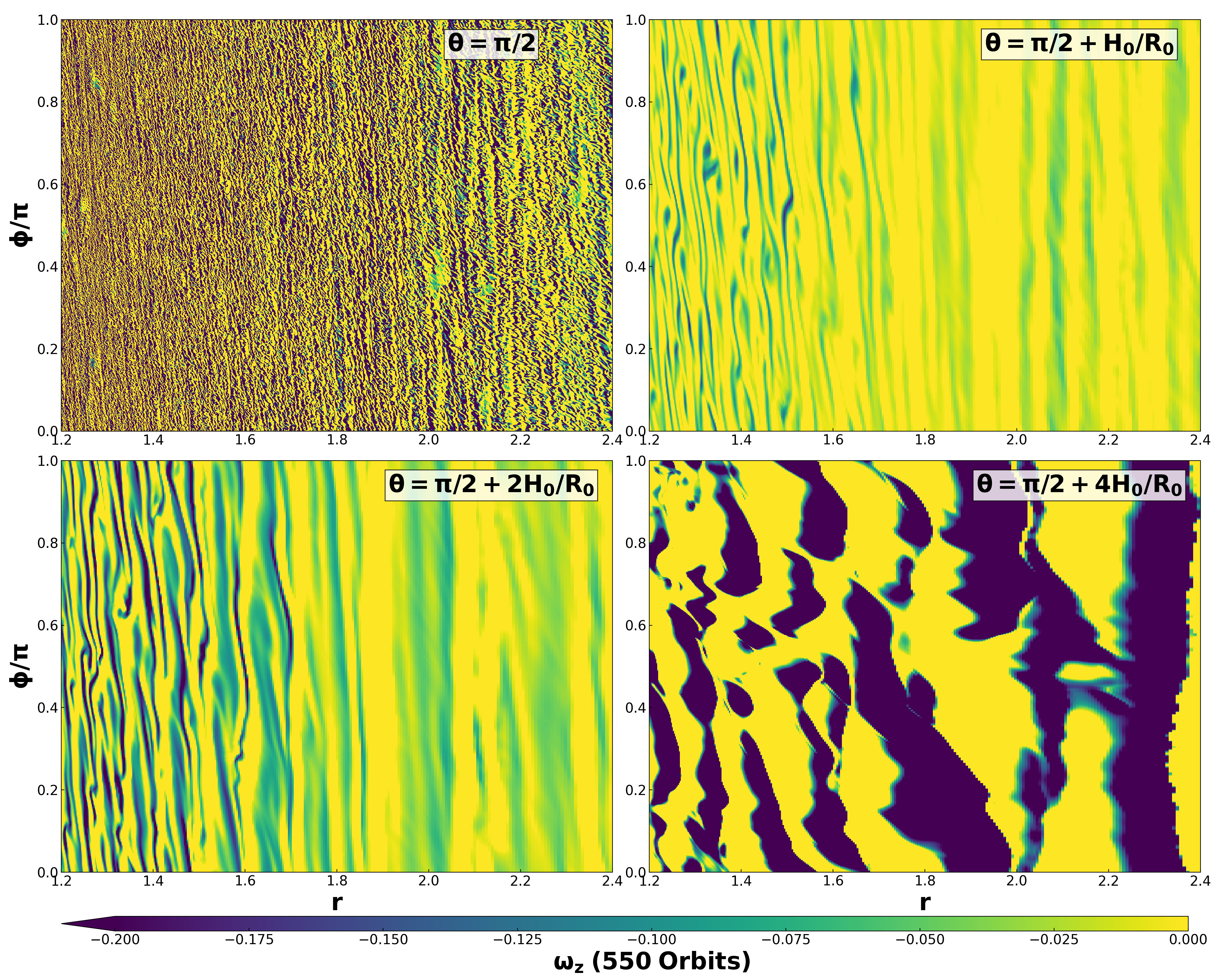}
\caption{The vertical vorticity $\omega_z$ at various $\theta-$angles is shown at the snapshot of 550 orbits for the ``3D-FIs'' model, similar to Figure 12 in HB25. We select $\theta = \pi/2$ ($z = 0$, midplane), $\pi/2 + H_0/R_0$, $\pi/2 + 2H_0/R_0$, and $\pi/2 + 4H_0/R_0$, where $H_0/R_0 = 0.08$ in this simulation. Note that the mesh level at $\theta = \pi/2$ is 3, while the mesh levels at other $\theta-$angles are 0 (root-grid level).}
\label{fig:Vorticity_z_3D_AMR}
\end{figure*}

\begin{figure*}[htp]
\centering
\includegraphics[scale=0.39]{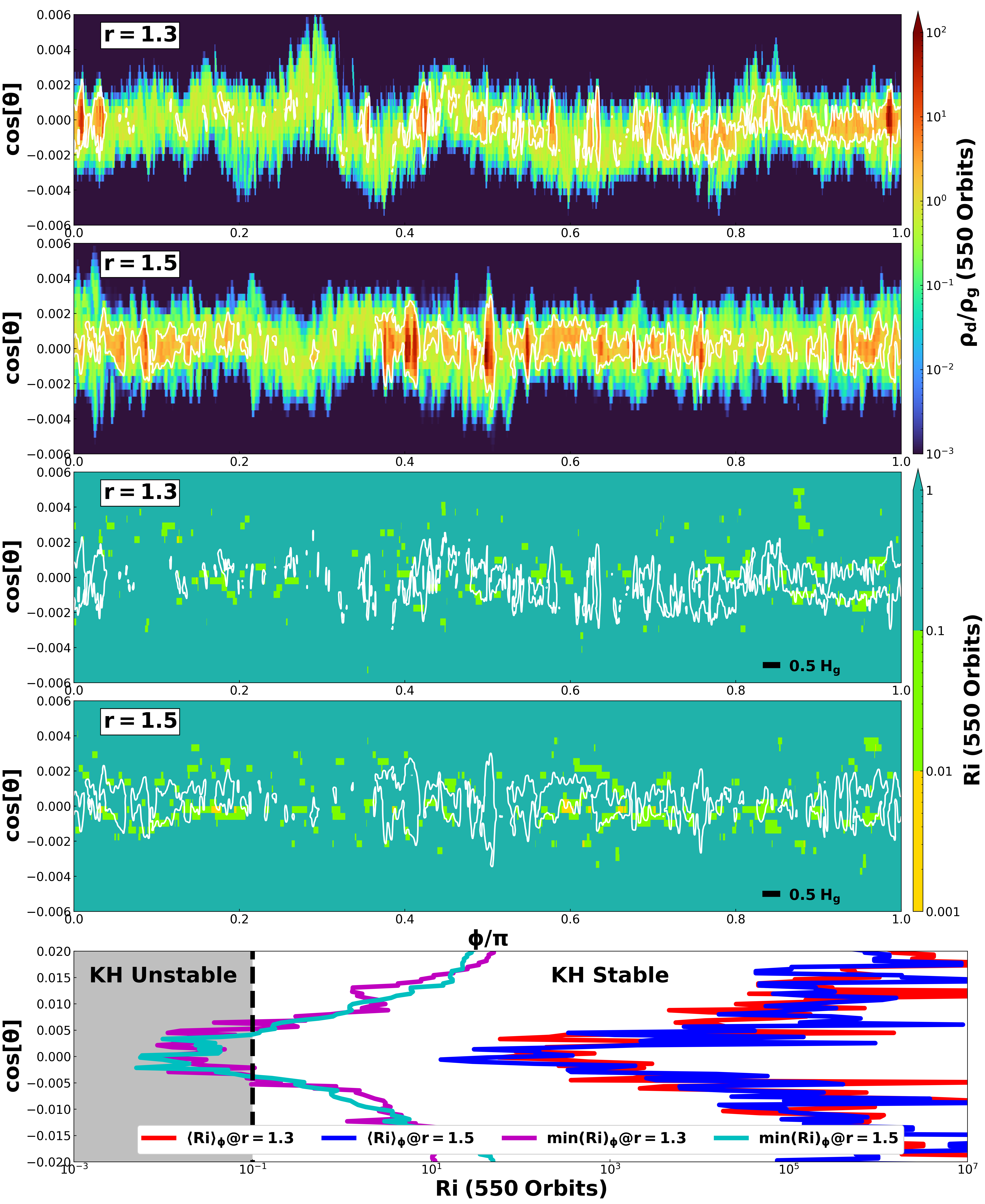}
\caption{The top two panels display the radial profiles of the dust-gas density ratio for the ``3D-FIs'' model at 550 orbits, taken at $r = 1.3$ and $1.5$. The middle two panels show the corresponding radial-azimuthal profiles of the Richardson number $Ri$ (Equation~\ref{eq:Ri_phi}), where we apply a moving-average with window width $0.5\;H_\text{g}$ along the azimuthal direction (indicated by the black solid segment in the bottom-right corners). White contours in the top four panels outline regions where the dust-gas density ratio exceeds unity. The bottom panel presents vertical profiles of the azimuthally averaged values of $Ri$ and the minimum $Ri$ from the moving average at the two radial locations.}
\label{fig:3D_dust_ratio_Ri_phi_3D_AMR}
\end{figure*}

In this subsection, we analyze in further detail the dust dynamics and the properties of dust clumping. In addition to the interplay between VSI and (dusty) RWI as studied in HB25, we highlight how the picture is enriched by the SI.

We start by looking at the horizontal and vertical distributions of the dust-gas density ratio, as shown in Figure~\ref{fig:3D_dust_ratio_3D_AMR} at the $\phi = \pi/2$ and $\theta = \pi/2$ ($z = 0$, midplane) slices over different time. From the horizontal distributions, we see that as the grid gets refined, we clearly identify the formation of dust clumps. Such clumps become much more densely populated towards higher resolution, and the individual clumps get more and more compact. From the vertical distribution, we see that most of the dust settles to very thin layers. There are regions where dust shows larger scale heights with corrugation patterns. They are associated with the transition between two neighboring inertial wave zones that propagate at the group velocity, as identified and discussed in Figure 10 of HB25. Interestingly, this region shows very little dust clumping even with much increased resolution.

The dust clumps are associated with dusty RWI vortices. In Figure~\ref{fig:3D_vorticity_ratio_3D_AMR}, we show zoomed-in views of the vertical vorticity and dust-gas density ratios at the midplane. Before activating mesh refinement (prior to 300 orbits), the dust-gas density ratio profiles resemble those of the ``3D-FB'' model. As refinement progresses, narrower dust rings form, which eventually break into smaller dust clumps, in accordance with the development of the dusty RWI~\citep{LiuBai2023}. Indeed, these dust rings and clumps are associated with regions of negative vertical vorticity, indicating that they correspond to dust concentrations within tiny dusty vortices (mild pressure bumps). We note that there is no one-to-one correspondence between gas vortices and dust clumps, are in general many more gas vortices. Moreover, only a small fraction ($1\%\sim2\%$) of dust clumps have density ratios exceeding the critical value, as highlighted with white contours.

The vortices show vertical structures. Figure~\ref{fig:Vorticity_z_3D_AMR} displays slices at constant $\theta$ of the vertical vorticity $\omega_z$ at 550 orbits. The midplane dust layer is the most densely populated with tiny vortices.
The vortex structures at $1\sim 2\;H_\text{g}$ above the midplane are similar, showing small vortices similar to those reported in HB25, while the vorticity increases over height. At higher altitudes ($\theta = \pi/2 + 4H_0/R_0$), density waves induced by VSI surface modes and larger vortices are observed.

Also shown in Figure~\ref{fig:3D_temperoal_evolution} are the cumulative distribution functions (CDFs) of the dust-gas density ratio exceeding specific threshold values, $P\left[ (\rho_\text{d}/\rho_\text{g}) > \left( \rho_\text{d}/\rho_\text{g} \right)_\text{threshold} \right]$, obtained by counting the number of cells. At level 0, the CDF agrees with the ``3D-FB'' model presented in HB25~\footnote{The CDF presented in HB25 extend further as it was measured at much later time. Our simulations also show this tendency, though we are not positioned to run much longer simulations given limited computing resources.}. With full three-level refinement, the tail of the CDF extends to $\rho_\text{d}/\rho_\text{g} \gtrsim 1000$. At this stage, the ratio of the mass of dust clumps with $(\rho_\text{d}/\rho_\text{g}) > (\rho_\text{d}/\rho_\text{g})_\text{crit}$ to the total dust mass is approximately $M_\text{clumps,total}/M_\text{dust,total} \sim 0.02 \text{--} 0.03$. Each clump contains roughly $10^{-6} \sim 10^{-5}$ of the total dust mass. Despite of reaching very high resolution sufficient to resolve the SI, we do not find sign of convergence of the CDF with increasing resolution. Note that even for pure SI, convergence in dust clumping behavior is known to require very high resolution, especially in 3D. Our study represents a step forward as the first study to investigate dust clumping under a self-consistent turbulent environment in a 3D global setup. Our results here mainly suggest that dust clumping and planetesimal formation can be more efficient than quoted above. Our results also call for independent verification from particle-based code as they may show different convergence properties in certain problems~\citep{Krapp2019,HuangBai2022}.

Based on the analysis above, we now discuss the role of SI on dust clumping. We first note that zonal flows from the VSI are not strong enough to overcome the background pressure gradient in the PPD (as discussed in HB25), thus the condition is favorable to promote the SI and dust clumping~\citep[e.g.][]{Bai2010dynamics}. From HB25, with low resolution, we also found that dust feedback already substantially enhances dust concentration. On the other hand, the association of dust clumps with vortices is suggestive of the dusty RWI~\citep{LiuBai2023}, although existing studies assume a pressure maxima and are only two-dimensional. We thus favor the interpretation of a combination of the SI and dusty RWI operating under such conditions. Finally, given that vortices are all very small and get smaller with increasing resolution, we do not expect to see the operation of the SI within vortices with migrating dust.

\subsection{Kelvin-Helmholtz Instability (KHI)}~\label{subsubsec:dustII}

With dust settling into a thin layer in the disk midplane, the gas azimuthal velocity is accelerated towards Keplerian by the strong dust feedback, while the gas above and below the dust layer is largely unaffected and still moves at non-Keplerian speed by the global (usually negative) pressure gradient. This gas azimuthal velocity transition leads to strong vertical shear in $v_{\text{g},\phi}$ and can be subject to the Kelvin-Helmholtz instability (KHI), which develops at the azimuthal-vertical ($\phi-z$) plane~\citep{Sekiya1998,YoudinShu2002,Chiang2008,LeeChiang2010a,LeeChiang2010b,Bai2010dynamics,Gerbig2020,SenguptaUmurhan2023}. One useful criterion for the onset of KHI is the Richardson number $Ri$, which can be defined as~\citep{Chiang2008,Bai2010dynamics}:
\begin{equation}
  Ri \equiv -\frac{GMz}{\rho_\text{eff} r^3}\frac{\partial \rho_\text{eff}/\partial z}{(\partial v_{\text{g},\phi}/\partial z)^2}
\label{eq:Ri_phi}
\end{equation}
For tightly coupled dust (with a Stokes number $St \lesssim 0.1$), the effective density is given by $\rho_{\text{eff}} = \rho_{\text{d}} + \rho_{\text{g}}$. In the absence of rotation, it is known that the KHI develops for $Ri < 0.25$~\citep{Cuzzi1993,Sekiya1998,YoudinShu2002,Johansen2006,LeeChiang2010a,LeeChiang2010b}. When considering the stabilizing effects from the Coriolis force,~\cite{Chiang2008} suggested that empirically, the critical value of $Ri$ is about $0.1$, which we adopt here.

To assess the potential activation of the KHI, we perform both a visual inspection of the dust layer at specific radial locations ($r = 1.3, 1.5$) based on data at 550 orbits and a more quantitative analysis of the Richardson number ($Ri$) profile at the corresponding positions. The results are presented in Figure~\ref{fig:3D_dust_ratio_Ri_phi_3D_AMR}.

The dust layer does exhibit distinct wave-like patterns in the azimuthal-vertical ($\phi-z$) plane at different radii, reminiscent of the activation of the KHI. When we compute the Richardson number profile by averaging data over the entire azimuthal domain, we observe that the Richardson number remains large ($Ri \gg 0.1$) beyond the dust layer, as expected. While it rapidly decreases toward the dust layer, the values remain well above $0.1$ (see the bottom panel of Figure~\ref{fig:3D_dust_ratio_Ri_phi_3D_AMR}). However, noticing that dust across the azimuth do not necessarily communicate, it is more reasonable to measure the {\it local} Richardson number, through a moving window of finite width along the $\phi$ direction. When taking the window size to be $0.5\;H_\text{g}$, which is more than 10 times the mean dust layer thickness, we see that the Richardson number can go below the critical value of $0.1$ (the green and yellow regions). Interestingly, most such regions coincide with the edges of the white contours, where the dust-gas density ratio is unity. Such contours can be considered as a marker for dust-gas interface, suggesting that the KHI can indeed be triggered near the such interfaces.

Interestingly, the dust density at the dust-gas interface exhibiting $Ri < 0.1$ is systematically lower ($\rho_\text{d}/\rho_\text{g} \lesssim 1$), and no dust clumps are observed in these regions. To further elucidate the role of KHI, we notice from Figure~~\ref{fig:3D_space_time_3D_AMR} that when averaged over azimuth, the dust ``rings'' persist over hundreds of orbits. On the other hand, dust in these narrow rings are spread into many small clumps within individual dusty vortices. Further examination from Figure~\ref{fig:3D_vorticity_ratio_3D_AMR} and neighboring snapshots reveals that the azimuthal dust distribution varies substantially from snapshot to snapshot (separated by one innermost orbit). This suggests that many of the clumps likely form and dissolve in transient, potentially due to the stirring (at least in part by) the KHI, although more detailed investigation is beyond the scope of this work.

\section{Conclusions and Discussion}~\label{sec:summary}

In this work, we conduct the first 3D global simulation of dust dynamics in PPDs subject to VSI with sufficient resolution to resolve the SI. With an initial dust-gas mass ratio of 0.01, we observe strong dust clumping and the complex interplay among the VSI, RWI, SI and KHI.

\begin{itemize}
  \item Among the four instabilities, the VSI sets the bulk level of turbulence and governs the overall gas dynamics, despite that the corrugation mode is suppressed, leaving the breathing mode the dominates dust stirring.
  \item The VSI generates weak zonal flows that do not necessarily form pressure maxima, allowing the SI to develop. Meanwhile, zonal flows facilitate the development of dusty RWI, leading to the formation of small dust clumps inside RWI vortices. Dust clumping is likely the outcome of both instabilities.
  \item The midplane layer is characterized by densely populated small gas vortices, though only a small fraction ($1\% \sim 2 \%$) of such vortices host dust clumps exceeded the Roche density.
  \item With increasing resolution, more dust participate in dust clumping, forming higher-density clumps, that do not show evidence of numerical convergence. About $2\% \sim 3\%$ of dust reside in regions clumps subject to gravitational collapse at our highest resolution.
  \item The dust layer is further affected by the KHI, which develops in localized azimuthal regions. It likely produces wave-like structures at the dust-gas interface along the azimuth, and contribute to temporary disruption of dust clumps.
\end{itemize}

Our work demonstrated that when accounting for dust feedback, dust clumping, likely triggered by the SI, can persist on top of 3D VSI, RWI as well as KHI turbulence. It suggests that under realistic environment, turbulence is not necessarily a diffusive and obstructive process for dust concentration, and dust back-reaction plays a crucial role assisting dust clumping. On the other hand, recall that dust clumping under the MRI turbulence likely occurs only in pressure maxima in addition to dust back-reaction~\citep{XuBai2022}, a condition different from what we observe here. Altogether, these works reveal the rich physics in dust-gas interaction and highlight the importance of self-consistent treatment of the gas turbulent environment, which is essential for a comprehensive understanding of dust dynamics and planetesimal formation.

Due to limited computational resources, we conducted only one single 3D global simulation in this paper with a set of parameters that we consider typical for general PPDs. Further investigations should explore more parameters such as dust size and abundance, as well as variations in the disk model.

We made several simplifying assumptions in terms of the physics incorporated, leaving room for future improvements. First, we simplify thermodynamics by fixing disk temperature with a locally isothermal equation of state.  More self-consistent treatment of radiation transport can alter the dominant mode of the VSI ~\citep{MelonFuksman2024,ZhangZhu2024}, which in the meantime may also trigger the convective overstability~\citep{Klahr2024}.  Intermediate cooling rates could also affect SI growth rates~\citep{LehmannLin2023} and the decay of RWI vortices ~\citep{FungOno2021}.  Second, we did not incorporate magnetic field, which likely dominate disk angular momentum transport by launching MHD winds, together with the MRI turbulence~\citep{CuiBai2021}. The MRI can co-exist with the VSI depending on disk ionization~\citep{CuiBai2022}. The disk wind configuration is subject to magnetic flux concentration to drive strong zonal flows~\citep{BaiStone2014,Suriano2018} and generate the RWI on its own~\citep{Hsu2024}. Third, self-gravity was neglected, but it becomes critical for understanding the initial mass function of planetesimals ~\citep{Johansen2015,Simon2016}. Finally, this study considered a single dust species, but dust size distribution can significantly affect disks cooling and hence the VSI turbulence ~\citep{Fukuhara2024}, as well as dust trapping in RWI vortices~\citep{LiLi2020Coagulation}, and potentially the properties of the SI as well~\citep{Krapp2019,YangZhu2021SI,ZhuYang2021,Matthijsse2025}. It should also be noted that dust coagulation can proceed efficiently in dust clumps, promoting further dust clumping and planetesimal formation~\citep{Tominaga2023,HoLiLi2024}. All of these effects are potentially fruitful avenues for future research.

\nolinenumbers
\section*{Acknowledgments}
  We thank Orkan Umurhan, Andrew Youdin, Hui Li, Eugene Chiang, Min-Kai Lin, Chao-Chin Yang, Shengtai Li, Can Cui, Rixin Li and Shangjia Zhang for helpful discussions. This work is supported by the National Science Foundation of China under grant No. 12233004, 12325304, 12342501, and the China Manned Space Project with NO. CMS-CSST-2021-B09. P. Huang acknowledge the Canadian Grant NSERC ALLRP 577027-22.

\appendix
\section{Theoretical Background}~\label{app:background}

\subsection{Vertical Shear Instability (VSI)}~\label{subsec:VSI}
In PPDs, the density profile is primarily set by angular momentum transport and stellar gravity, while the temperature profile is predominantly regulated by stellar irradiation~\citep{ChiangGoldreich1997}. Notably, the contours of constant density and pressure are generally not parallel, making PPDs essentially non-barotropic ($\nabla \rho_\text{g} \times \nabla P \ne 0$). This leads to the presence of weak vertical shear~\citep{BarkerLatter2015}:

\begin{equation}
  2 R \Omega_{\text{g}} \frac{\partial \Omega_\text{g}}{\partial z} = - \frac{\left(\nabla \rho_\text{g} \times \nabla P\right)}{\rho_\text{g}^2} \cdot \hat{\phi}
\label{eq:vert_shear}
\end{equation}

When subjected to rapid thermal cooling (the thermal relaxation timescale is much smaller than Keplerian dynamic timescale, $t_\text{cool} \ll \Omega_\text{K}^{-1} $), the strong stabilizing vertical buoyancy within disks is diminished~\citep{LinYoudin2015}.
This triggers the vertical shear instability~\cite[VSI,][]{UrpinBrandenburg1998,Urpin2003,ArltUrpin2004,Nelson2013}, which is an axisymmetric ($R-z$ plane) and centrifugal instability analogous to the Goldreich-Schubert-Fricke (GSF) instability in differentially rotating stars~\citep{GoldreichSchubert1967,Fricke1968}.

The VSI taps the free energy from vertical shear in disks
and generates modest and anisotropic turbulence characterized by the ``body modes''. They result from destabilization of the inertial waves and are usually dominated by strong vertical oscillation, which levitate dust particles~\citep{StollKley2016,Flock2020,Dullemond2022}. Furthermore, VSI can generate weak zonal flows and vortices from secondary instabilities (RWI), facilitating dust trapping/concentration~\citep{Richard2016,Flock2017a,MangerKlahr2018,Manger2020,Flock2020}.

\subsection{Streaming Instability (SI)}~\label{subsec:SI}

In PPDs, dust particles undergo radial drift towards higher gas pressure~\citep{Weidenschilling1977Aerodynamics}.
When considering the dust back-reaction to the gas, this motion becomes linearly unstable to axisymmetric perturbations, leading to the streaming instability~\cite[SI,][]{GoodmanPindor2000,YoudinGoodman2005,YoudinJohansen2007}.
It can also be interpreted as a resonant drag instability~\cite[RDI,][]{HopkinsSquire18,SquireHopkins2018,SquireHopkins2020},
resulting from the resonance between the dust streaming motion (drift velocity $\mathbf{w}$) and gas epicyclic motion (epicyclic frequency $\kappa$), i.e., $\mathbf{w} \cdot \hat{k} = \kappa$, where $\hat{k}$ is the projected wave vector. The free energy of SI comes from the streaming motion due to radial pressure gradient, characterized by
\begin{equation}
  \eta \equiv  \frac{1}{2} \left(\frac{c_\text{s}}{v_\text{K}}\right)^2 \frac{d \ln P}{d \ln R}
\label{eq:presgra_eta}
\end{equation}
and it is often non-dimensionized by $\Pi\equiv\eta v_\text{K}/c_\text{s}$, where $c_\text{s}$ and $v_\text{K}$ are the gas sound speed and Keplerian speed.

In the non-linear regime, gas drag, dust feedback, and the Coriolis force can create a positive loop that amplifies dust density fluctuations, leading to strong dust overdensities~\citep{JohansenYoudin2007,SquireHopkins2018,Magnan2024}. Considering vertical stratification, the SI can yield significant dust clumping depending on the dust abundance, size distribution and radial pressure gradient~\citep{Bai2010pressure,Bai2010dynamics,CarreraJohansen2015,LiYoudin2021,LimSimon2024b,OstertagFlock2025}, and when further incorporating self-gravity, such dust clumps can further collapse to directly form planetesimals~\citep{JohansenOishi2007,Simon2016,LiYoudin2018}.

The interplay between VSI and SI has recently been investigated in ~\cite{SchaferJohansen2020,SchaferJohansen2022,Schafer2025}, as well as our work (HB25), in a 2D axisymmetric setting. It was found that SI exhibits the capacity to generate more robust dust clumping when compared to scenarios where SI acts alone.

\subsection{Rossby Wave Instability (RWI)}~\label{subsec:RWI}

The Rossby wave instability is a local non-axisymmetric instability~\cite[RWI,][]{LovelaceLi1999,LiFinn2000,Ono2016}, which occurs in regions where there is a local maximum of the key function $\mathscr{L}$ known as ``Entropy Modified Inversed Vortensity'':
\begin{equation}
  \mathscr{L} \equiv  \frac{\Sigma_\text{g} s^{2/\gamma}}{2 \left(\nabla \times \mathbf{v}_\text{g}\right)\cdot \hat{z}}
\label{eq:Lfunc}
\end{equation}
where $\Sigma_\text{g}$, $\gamma$, $\mathbf{v}_\text{g}$ and $s$ are gas surface density, adiabatic index, gas velocities and specific gas entropy, respectively. The specific entropy is defined as $s\equiv p_\text{g}/\Sigma_\text{g}^{\gamma}$, and $p_\text{g}$ is the vertical integrated pressure. A local maximum of $\mathscr{L}$ could occur in locations characterized by anomalous radial shear at orbital frequency $\partial \Omega_\text{g}/\partial R$ compared to background Keplerian shear, typically associated with strong radial pressure variations. Examples of such locations include gap edges created by planets~\citep{deVal-Borro2007,LinPapaloizou2011,ZhuStone2014}, zonal flows~\citep{Johansen2009,Richard2016}, mass infall~\citep{Bae2015,Kuznetsova2022}, boundary layer~\citep{FuHuangYu2023} and dusty rings edges~\citep{FuLi2014,PierensLin2019,YangZhu2020,HuangLiIsella2020,HsiehLin2020,LiuBai2023,ChanPaardekooper2024}. Since the anomalous radial shear is important, RWI can be considered as a special Kelvin-Helmholtz type instability acting on the horizontal ($R-\phi$) plane for accretion disks.

The non-linear development of the RWI in disks usually leads to the production of multiple anticyclonic vortices, which may eventually merge into one big vortex~\citep{LiColgate2001,Meheut2012Vortices3D,Meheut2012Dust,Meheut2012RWI,Ono2018}. The vortices can launch density waves~\citep{Bodo2005,Meheut2010,HuangDong2019,Ma2025}, which lead to their migration~\citep{LiColgate2001,Paardekooper2010,Ono2018}. A most attractive property of RWI vortices is that their anticyclonic nature are characterized by a pressure maxima at the center, thus are expected to be able to concentrate dust particles and facilitate planetesimal formation~\citep{BargeSommeria1995}. Recent observations made by the Atacama Large Millimeter/submillimeter Array (ALMA) indicate that azimuthal asymmetries in the dust continuum are not uncommon, and may be attributed to the trapping of dust within RWI vortices~\citep{vanderMarel2013,vanderMarel2016,Dong2018MWC758,Yang2023}.

\vspace{5mm}

\software{Athena++~\citep{Stone2020,HuangBai2022}}

\bibliographystyle{aasjournal}
\bibliography{references}{}

\begin{thebibliography}{}
\expandafter\ifx\csname natexlab\endcsname\relax\def\natexlab#1{#1}\fi
\providecommand{\url}[1]{\href{#1}{#1}}
\providecommand{\dodoi}[1]{doi:~\href{http://doi.org/#1}{\nolinkurl{#1}}}
\providecommand{\doeprint}[1]{\href{http://ascl.net/#1}{\nolinkurl{http://ascl.net/#1}}}
\providecommand{\doarXiv}[1]{\href{https://arxiv.org/abs/#1}{\nolinkurl{https://arxiv.org/abs/#1}}}

\bibitem[{{Arlt} \& {Urpin}(2004)}]{ArltUrpin2004}
{Arlt}, R., \& {Urpin}, V. 2004, \aap, 426, 755,
  \dodoi{10.1051/0004-6361:20035896}

\bibitem[{{Bae} {et~al.}(2015){Bae}, {Hartmann}, \& {Zhu}}]{Bae2015}
{Bae}, J., {Hartmann}, L., \& {Zhu}, Z. 2015, \apj, 805, 15,
  \dodoi{10.1088/0004-637X/805/1/15}

\bibitem[{{Bae} {et~al.}(2023){Bae}, {Isella}, {Zhu}, {Martin}, {Okuzumi}, \&
  {Suriano}}]{Bae2023}
{Bae}, J., {Isella}, A., {Zhu}, Z., {et~al.} 2023, in Astronomical Society of
  the Pacific Conference Series, Vol. 534, Protostars and Planets VII, ed.
  S.~{Inutsuka}, Y.~{Aikawa}, T.~{Muto}, K.~{Tomida}, \& M.~{Tamura}, 423,
  \dodoi{10.48550/arXiv.2210.13314}

\bibitem[{{Bai} \& {Stone}(2010{\natexlab{a}})}]{Bai2010dynamics}
{Bai}, X.-N., \& {Stone}, J.~M. 2010{\natexlab{a}}, \apj, 722, 1437,
  \dodoi{10.1088/0004-637X/722/2/1437}

\bibitem[{{Bai} \& {Stone}(2010{\natexlab{b}})}]{Bai2010pressure}
---. 2010{\natexlab{b}}, \apjl, 722, L220, \dodoi{10.1088/2041-8205/722/2/L220}

\bibitem[{{Bai} \& {Stone}(2011)}]{BaiStone2011}
---. 2011, \apj, 736, 144, \dodoi{10.1088/0004-637X/736/2/144}

\bibitem[{{Bai} \& {Stone}(2014)}]{BaiStone2014}
---. 2014, \apj, 796, 31, \dodoi{10.1088/0004-637X/796/1/31}

\bibitem[{{Balbus} \& {Hawley}(1998)}]{BalbusHawley1998}
{Balbus}, S.~A., \& {Hawley}, J.~F. 1998, Reviews of Modern Physics, 70, 1,
  \dodoi{10.1103/RevModPhys.70.1}

\bibitem[{{Barge} \& {Sommeria}(1995)}]{BargeSommeria1995}
{Barge}, P., \& {Sommeria}, J. 1995, \aap, 295, L1.
\newblock \doarXiv{astro-ph/9501050}

\bibitem[{{Barker} \& {Latter}(2015)}]{BarkerLatter2015}
{Barker}, A.~J., \& {Latter}, H.~N. 2015, \mnras, 450, 21,
  \dodoi{10.1093/mnras/stv640}

\bibitem[{{Bodo} {et~al.}(2005){Bodo}, {Chagelishvili}, {Murante}, {Tevzadze},
  {Rossi}, \& {Ferrari}}]{Bodo2005}
{Bodo}, G., {Chagelishvili}, G., {Murante}, G., {et~al.} 2005, \aap, 437, 9,
  \dodoi{10.1051/0004-6361:20041046}

\bibitem[{{Carrera} {et~al.}(2015){Carrera}, {Johansen}, \&
  {Davies}}]{CarreraJohansen2015}
{Carrera}, D., {Johansen}, A., \& {Davies}, M.~B. 2015, \aap, 579, A43,
  \dodoi{10.1051/0004-6361/201425120}

\bibitem[{{Chan} \& {Paardekooper}(2024)}]{ChanPaardekooper2024}
{Chan}, K., \& {Paardekooper}, S.-J. 2024, \mnras, 528, 5904,
  \dodoi{10.1093/mnras/stae089}

\bibitem[{{Chen} \& {Lin}(2020)}]{ChenLin2020}
{Chen}, K., \& {Lin}, M.-K. 2020, \apj, 891, 132,
  \dodoi{10.3847/1538-4357/ab76ca}

\bibitem[{{Chiang}(2008)}]{Chiang2008}
{Chiang}, E. 2008, \apj, 675, 1549, \dodoi{10.1086/527354}

\bibitem[{{Chiang} \& {Youdin}(2010)}]{ChiangYoudin2010}
{Chiang}, E., \& {Youdin}, A.~N. 2010, Annual Review of Earth and Planetary
  Sciences, 38, 493, \dodoi{10.1146/annurev-earth-040809-152513}

\bibitem[{{Chiang} \& {Goldreich}(1997)}]{ChiangGoldreich1997}
{Chiang}, E.~I., \& {Goldreich}, P. 1997, \apj, 490, 368,
  \dodoi{10.1086/304869}

\bibitem[{{Cui} \& {Bai}(2021)}]{CuiBai2021}
{Cui}, C., \& {Bai}, X.-N. 2021, \mnras, 507, 1106,
  \dodoi{10.1093/mnras/stab2220}

\bibitem[{{Cui} \& {Bai}(2022)}]{CuiBai2022}
---. 2022, \mnras, 516, 4660, \dodoi{10.1093/mnras/stac2580}

\bibitem[{{Cuzzi} {et~al.}(1993){Cuzzi}, {Dobrovolskis}, \&
  {Champney}}]{Cuzzi1993}
{Cuzzi}, J.~N., {Dobrovolskis}, A.~R., \& {Champney}, J.~M. 1993, \icarus, 106,
  102, \dodoi{10.1006/icar.1993.1161}

\bibitem[{{de Val-Borro} {et~al.}(2007){de Val-Borro}, {Artymowicz},
  {D'Angelo}, \& {Peplinski}}]{deVal-Borro2007}
{de Val-Borro}, M., {Artymowicz}, P., {D'Angelo}, G., \& {Peplinski}, A. 2007,
  \aap, 471, 1043, \dodoi{10.1051/0004-6361:20077169}

\bibitem[{{Dong} {et~al.}(2018){Dong}, {Liu}, {Eisner}, {Andrews}, {Fung},
  {Zhu}, {Chiang}, {Hashimoto}, {Liu}, {Casassus}, {Esposito}, {Hasegawa},
  {Muto}, {Pavlyuchenkov}, {Wilner}, {Akiyama}, {Tamura}, \&
  {Wisniewski}}]{Dong2018MWC758}
{Dong}, R., {Liu}, S.-y., {Eisner}, J., {et~al.} 2018, \apj, 860, 124,
  \dodoi{10.3847/1538-4357/aac6cb}

\bibitem[{{Dullemond} {et~al.}(2022){Dullemond}, {Ziampras}, {Ostertag}, \&
  {Dominik}}]{Dullemond2022}
{Dullemond}, C.~P., {Ziampras}, A., {Ostertag}, D., \& {Dominik}, C. 2022,
  \aap, 668, A105, \dodoi{10.1051/0004-6361/202244218}

\bibitem[{{Flock} {et~al.}(2017){Flock}, {Nelson}, {Turner}, {Bertrang},
  {Carrasco-Gonz{\'a}lez}, {Henning}, {Lyra}, \& {Teague}}]{Flock2017a}
{Flock}, M., {Nelson}, R.~P., {Turner}, N.~J., {et~al.} 2017, \apj, 850, 131,
  \dodoi{10.3847/1538-4357/aa943f}

\bibitem[{{Flock} {et~al.}(2020){Flock}, {Turner}, {Nelson}, {Lyra}, {Manger},
  \& {Klahr}}]{Flock2020}
{Flock}, M., {Turner}, N.~J., {Nelson}, R.~P., {et~al.} 2020, \apj, 897, 155,
  \dodoi{10.3847/1538-4357/ab9641}

\bibitem[{{Fricke}(1968)}]{Fricke1968}
{Fricke}, K. 1968, \zap, 68, 317

\bibitem[{{Fu} {et~al.}(2014){Fu}, {Li}, {Lubow}, {Li}, \& {Liang}}]{FuLi2014}
{Fu}, W., {Li}, H., {Lubow}, S., {Li}, S., \& {Liang}, E. 2014, \apjl, 795,
  L39, \dodoi{10.1088/2041-8205/795/2/L39}

\bibitem[{{Fu} {et~al.}(2023){Fu}, {Huang}, \& {Yu}}]{FuHuangYu2023}
{Fu}, Z., {Huang}, S., \& {Yu}, C. 2023, \apj, 945, 165,
  \dodoi{10.3847/1538-4357/acac9c}

\bibitem[{{Fukuhara} \& {Okuzumi}(2024)}]{Fukuhara2024}
{Fukuhara}, Y., \& {Okuzumi}, S. 2024, arXiv e-prints, arXiv:2404.15780,
  \dodoi{10.48550/arXiv.2404.15780}

\bibitem[{{Fung} \& {Ono}(2021)}]{FungOno2021}
{Fung}, J., \& {Ono}, T. 2021, \apj, 922, 13, \dodoi{10.3847/1538-4357/ac1d4e}

\bibitem[{{Gerbig} {et~al.}(2020){Gerbig}, {Murray-Clay}, {Klahr}, \&
  {Baehr}}]{Gerbig2020}
{Gerbig}, K., {Murray-Clay}, R.~A., {Klahr}, H., \& {Baehr}, H. 2020, \apj,
  895, 91, \dodoi{10.3847/1538-4357/ab8d37}

\bibitem[{{Goldreich} \& {Schubert}(1967)}]{GoldreichSchubert1967}
{Goldreich}, P., \& {Schubert}, G. 1967, \apj, 150, 571, \dodoi{10.1086/149360}

\bibitem[{{Gole} {et~al.}(2020){Gole}, {Simon}, {Li}, {Youdin}, \&
  {Armitage}}]{Gole2020}
{Gole}, D.~A., {Simon}, J.~B., {Li}, R., {Youdin}, A.~N., \& {Armitage}, P.~J.
  2020, \apj, 904, 132, \dodoi{10.3847/1538-4357/abc334}

\bibitem[{{Goodman} \& {Pindor}(2000)}]{GoodmanPindor2000}
{Goodman}, J., \& {Pindor}, B. 2000, \icarus, 148, 537,
  \dodoi{10.1006/icar.2000.6467}

\bibitem[{{Ho} {et~al.}(2024){Ho}, {Li}, \& {Li}}]{HoLiLi2024}
{Ho}, K.~W., {Li}, H., \& {Li}, S. 2024, \apjl, 975, L34,
  \dodoi{10.3847/2041-8213/ad8655}

\bibitem[{{Hopkins} \& {Squire}(2018)}]{HopkinsSquire18}
{Hopkins}, P.~F., \& {Squire}, J. 2018, \mnras, 479, 4681,
  \dodoi{10.1093/mnras/sty1604}

\bibitem[{{Hsieh} \& {Lin}(2020)}]{HsiehLin2020}
{Hsieh}, H.-F., \& {Lin}, M.-K. 2020, \mnras, 497, 2425,
  \dodoi{10.1093/mnras/staa2115}

\bibitem[{{Hsu} {et~al.}(2024){Hsu}, {Li}, {Tu}, {Hu}, \& {Lin}}]{Hsu2024}
{Hsu}, C.-Y., {Li}, Z.-Y., {Tu}, Y., {Hu}, X., \& {Lin}, M.-K. 2024, \mnras,
  533, 2980, \dodoi{10.1093/mnras/stae1986}

\bibitem[{{Huang} \& {Bai}(2022)}]{HuangBai2022}
{Huang}, P., \& {Bai}, X.-N. 2022, \apjs, 262, 11,
  \dodoi{10.3847/1538-4365/ac76cb}

\bibitem[{{Huang} \& {Bai}(2025)}]{HuangBai2025I}
---. 2025, arXiv preprint arXiv:2503.01656

\bibitem[{{Huang} {et~al.}(2019){Huang}, {Dong}, {Li}, {Li}, \&
  {Ji}}]{HuangDong2019}
{Huang}, P., {Dong}, R., {Li}, H., {Li}, S., \& {Ji}, J. 2019, \apjl, 883, L39,
  \dodoi{10.3847/2041-8213/ab40c4}

\bibitem[{{Huang} {et~al.}(2020){Huang}, {Li}, {Isella}, {Miranda}, {Li}, \&
  {Ji}}]{HuangLiIsella2020}
{Huang}, P., {Li}, H., {Isella}, A., {et~al.} 2020, \apj, 893, 89,
  \dodoi{10.3847/1538-4357/ab8199}

\bibitem[{{Johansen} {et~al.}(2006){Johansen}, {Henning}, \&
  {Klahr}}]{Johansen2006}
{Johansen}, A., {Henning}, T., \& {Klahr}, H. 2006, \apj, 643, 1219,
  \dodoi{10.1086/502968}

\bibitem[{{Johansen} {et~al.}(2015){Johansen}, {Mac Low}, {Lacerda}, \&
  {Bizzarro}}]{Johansen2015}
{Johansen}, A., {Mac Low}, M.-M., {Lacerda}, P., \& {Bizzarro}, M. 2015,
  Science Advances, 1, 1500109, \dodoi{10.1126/sciadv.1500109}

\bibitem[{{Johansen} {et~al.}(2007){Johansen}, {Oishi}, {Mac Low}, {Klahr},
  {Henning}, \& {Youdin}}]{JohansenOishi2007}
{Johansen}, A., {Oishi}, J.~S., {Mac Low}, M.-M., {et~al.} 2007, \nat, 448,
  1022, \dodoi{10.1038/nature06086}

\bibitem[{{Johansen} \& {Youdin}(2007)}]{JohansenYoudin2007}
{Johansen}, A., \& {Youdin}, A. 2007, \apj, 662, 627, \dodoi{10.1086/516730}

\bibitem[{{Johansen} {et~al.}(2009){Johansen}, {Youdin}, \&
  {Klahr}}]{Johansen2009}
{Johansen}, A., {Youdin}, A., \& {Klahr}, H. 2009, \apj, 697, 1269,
  \dodoi{10.1088/0004-637X/697/2/1269}

\bibitem[{{Klahr}(2024)}]{Klahr2024}
{Klahr}, H. 2024, arXiv e-prints, arXiv:2404.15933,
  \dodoi{10.48550/arXiv.2404.15933}

\bibitem[{{Krapp} {et~al.}(2019){Krapp}, {Ben{\'\i}tez-Llambay}, {Gressel}, \&
  {Pessah}}]{Krapp2019}
{Krapp}, L., {Ben{\'\i}tez-Llambay}, P., {Gressel}, O., \& {Pessah}, M.~E.
  2019, \apjl, 878, L30, \dodoi{10.3847/2041-8213/ab2596}

\bibitem[{{Kuznetsova} {et~al.}(2022){Kuznetsova}, {Bae}, {Hartmann}, \& {Mac
  Low}}]{Kuznetsova2022}
{Kuznetsova}, A., {Bae}, J., {Hartmann}, L., \& {Mac Low}, M.-M. 2022, \apj,
  928, 92, \dodoi{10.3847/1538-4357/ac54a8}

\bibitem[{{Lee} {et~al.}(2010{\natexlab{a}}){Lee}, {Chiang}, {Asay-Davis}, \&
  {Barranco}}]{LeeChiang2010a}
{Lee}, A.~T., {Chiang}, E., {Asay-Davis}, X., \& {Barranco}, J.
  2010{\natexlab{a}}, \apj, 718, 1367, \dodoi{10.1088/0004-637X/718/2/1367}

\bibitem[{{Lee} {et~al.}(2010{\natexlab{b}}){Lee}, {Chiang}, {Asay-Davis}, \&
  {Barranco}}]{LeeChiang2010b}
---. 2010{\natexlab{b}}, \apj, 725, 1938, \dodoi{10.1088/0004-637X/725/2/1938}

\bibitem[{{Lehmann} \& {Lin}(2023)}]{LehmannLin2023}
{Lehmann}, M., \& {Lin}, M.-K. 2023, \mnras, 522, 5892,
  \dodoi{10.1093/mnras/stad1349}

\bibitem[{{Li} {et~al.}(2001){Li}, {Colgate}, {Wendroff}, \&
  {Liska}}]{LiColgate2001}
{Li}, H., {Colgate}, S.~A., {Wendroff}, B., \& {Liska}, R. 2001, \apj, 551,
  874, \dodoi{10.1086/320241}

\bibitem[{{Li} {et~al.}(2000){Li}, {Finn}, {Lovelace}, \&
  {Colgate}}]{LiFinn2000}
{Li}, H., {Finn}, J.~M., {Lovelace}, R.~V.~E., \& {Colgate}, S.~A. 2000, \apj,
  533, 1023, \dodoi{10.1086/308693}

\bibitem[{{Li} \& {Youdin}(2021)}]{LiYoudin2021}
{Li}, R., \& {Youdin}, A.~N. 2021, \apj, 919, 107,
  \dodoi{10.3847/1538-4357/ac0e9f}

\bibitem[{{Li} {et~al.}(2018){Li}, {Youdin}, \& {Simon}}]{LiYoudin2018}
{Li}, R., {Youdin}, A.~N., \& {Simon}, J.~B. 2018, \apj, 862, 14,
  \dodoi{10.3847/1538-4357/aaca99}

\bibitem[{{Li} {et~al.}(2020){Li}, {Li}, {Li}, {Birnstiel}, {Dra{\.z}kowska},
  \& {Stammler}}]{LiLi2020Coagulation}
{Li}, Y.-P., {Li}, H., {Li}, S., {et~al.} 2020, \apjl, 892, L19,
  \dodoi{10.3847/2041-8213/ab7fb2}

\bibitem[{{Lim} {et~al.}(2024{\natexlab{a}}){Lim}, {Simon}, {Li}, {Carrera},
  {Baronett}, {Youdin}, {Lyra}, \& {Yang}}]{LimSimon2024b}
{Lim}, J., {Simon}, J.~B., {Li}, R., {et~al.} 2024{\natexlab{a}}, arXiv
  e-prints, arXiv:2410.17319, \dodoi{10.48550/arXiv.2410.17319}

\bibitem[{{Lim} {et~al.}(2024{\natexlab{b}}){Lim}, {Simon}, {Li}, {Armitage},
  {Carrera}, {Lyra}, {Rea}, {Yang}, \& {Youdin}}]{Lim2024}
---. 2024{\natexlab{b}}, \apj, 969, 130, \dodoi{10.3847/1538-4357/ad47a2}

\bibitem[{{Lin} \& {Papaloizou}(2011)}]{LinPapaloizou2011}
{Lin}, M.-K., \& {Papaloizou}, J. C.~B. 2011, \mnras, 415, 1426,
  \dodoi{10.1111/j.1365-2966.2011.18798.x}

\bibitem[{{Lin} \& {Youdin}(2015)}]{LinYoudin2015}
{Lin}, M.-K., \& {Youdin}, A.~N. 2015, \apj, 811, 17,
  \dodoi{10.1088/0004-637X/811/1/17}

\bibitem[{{Liu} \& {Bai}(2023)}]{LiuBai2023}
{Liu}, H., \& {Bai}, X.-N. 2023, \mnras, \dodoi{10.1093/mnras/stad2629}

\bibitem[{{Lovelace} {et~al.}(1999){Lovelace}, {Li}, {Colgate}, \&
  {Nelson}}]{LovelaceLi1999}
{Lovelace}, R.~V.~E., {Li}, H., {Colgate}, S.~A., \& {Nelson}, A.~F. 1999,
  \apj, 513, 805, \dodoi{10.1086/306900}

\bibitem[{{Ma} {et~al.}(2025){Ma}, {Huang}, {Yu}, \& {Dong}}]{Ma2025}
{Ma}, X., {Huang}, P., {Yu}, C., \& {Dong}, R. 2025, \apj, 979, 244,
  \dodoi{10.3847/1538-4357/ad9f2c}

\bibitem[{{Magnan} {et~al.}(2024){Magnan}, {Heinemann}, \&
  {Latter}}]{Magnan2024}
{Magnan}, N., {Heinemann}, T., \& {Latter}, H.~N. 2024, \mnras, 534, 3944,
  \dodoi{10.1093/mnras/stae1978}

\bibitem[{{Manger} \& {Klahr}(2018)}]{MangerKlahr2018}
{Manger}, N., \& {Klahr}, H. 2018, \mnras, 480, 2125,
  \dodoi{10.1093/mnras/sty1909}

\bibitem[{{Manger} {et~al.}(2020){Manger}, {Klahr}, {Kley}, \&
  {Flock}}]{Manger2020}
{Manger}, N., {Klahr}, H., {Kley}, W., \& {Flock}, M. 2020, \mnras, 499, 1841,
  \dodoi{10.1093/mnras/staa2943}

\bibitem[{{Matthijsse} {et~al.}(2025){Matthijsse}, {Aly}, \&
  {Paardekooper}}]{Matthijsse2025}
{Matthijsse}, J., {Aly}, H., \& {Paardekooper}, S.-J. 2025, arXiv e-prints,
  arXiv:2502.01752, \dodoi{10.48550/arXiv.2502.01752}

\bibitem[{{Meheut} {et~al.}(2010){Meheut}, {Casse}, {Varniere}, \&
  {Tagger}}]{Meheut2010}
{Meheut}, H., {Casse}, F., {Varniere}, P., \& {Tagger}, M. 2010, \aap, 516,
  A31, \dodoi{10.1051/0004-6361/201014000}

\bibitem[{{Meheut} {et~al.}(2012{\natexlab{a}}){Meheut}, {Keppens}, {Casse}, \&
  {Benz}}]{Meheut2012Vortices3D}
{Meheut}, H., {Keppens}, R., {Casse}, F., \& {Benz}, W. 2012{\natexlab{a}},
  \aap, 542, A9, \dodoi{10.1051/0004-6361/201118500}

\bibitem[{{Meheut} {et~al.}(2012{\natexlab{b}}){Meheut}, {Meliani}, {Varniere},
  \& {Benz}}]{Meheut2012Dust}
{Meheut}, H., {Meliani}, Z., {Varniere}, P., \& {Benz}, W. 2012{\natexlab{b}},
  \aap, 545, A134, \dodoi{10.1051/0004-6361/201219794}

\bibitem[{{Meheut} {et~al.}(2012{\natexlab{c}}){Meheut}, {Yu}, \&
  {Lai}}]{Meheut2012RWI}
{Meheut}, H., {Yu}, C., \& {Lai}, D. 2012{\natexlab{c}}, \mnras, 422, 2399,
  \dodoi{10.1111/j.1365-2966.2012.20789.x}

\bibitem[{{Melon Fuksman} {et~al.}(2024){Melon Fuksman}, {Flock}, \&
  {Klahr}}]{MelonFuksman2024}
{Melon Fuksman}, J.~D., {Flock}, M., \& {Klahr}, H. 2024, \aap, 682, A139,
  \dodoi{10.1051/0004-6361/202346554}

\bibitem[{{Nelson} {et~al.}(2013){Nelson}, {Gressel}, \&
  {Umurhan}}]{Nelson2013}
{Nelson}, R.~P., {Gressel}, O., \& {Umurhan}, O.~M. 2013, \mnras, 435, 2610,
  \dodoi{10.1093/mnras/stt1475}

\bibitem[{{Ono} {et~al.}(2016){Ono}, {Muto}, {Takeuchi}, \& {Nomura}}]{Ono2016}
{Ono}, T., {Muto}, T., {Takeuchi}, T., \& {Nomura}, H. 2016, \apj, 823, 84,
  \dodoi{10.3847/0004-637X/823/2/84}

\bibitem[{{Ono} {et~al.}(2018){Ono}, {Muto}, {Tomida}, \& {Zhu}}]{Ono2018}
{Ono}, T., {Muto}, T., {Tomida}, K., \& {Zhu}, Z. 2018, \apj, 864, 70,
  \dodoi{10.3847/1538-4357/aad54d}

\bibitem[{{Ostertag} \& {Flock}(2025)}]{OstertagFlock2025}
{Ostertag}, D., \& {Flock}, M. 2025, arXiv e-prints, arXiv:2501.18424,
  \dodoi{10.48550/arXiv.2501.18424}

\bibitem[{{Paardekooper} {et~al.}(2010){Paardekooper}, {Lesur}, \&
  {Papaloizou}}]{Paardekooper2010}
{Paardekooper}, S.-J., {Lesur}, G., \& {Papaloizou}, J. C.~B. 2010, \apj, 725,
  146, \dodoi{10.1088/0004-637X/725/1/146}

\bibitem[{{Pierens} {et~al.}(2019){Pierens}, {Lin}, \&
  {Raymond}}]{PierensLin2019}
{Pierens}, A., {Lin}, M.~K., \& {Raymond}, S.~N. 2019, \mnras, 488, 645,
  \dodoi{10.1093/mnras/stz1718}

\bibitem[{{Richard} {et~al.}(2016){Richard}, {Nelson}, \&
  {Umurhan}}]{Richard2016}
{Richard}, S., {Nelson}, R.~P., \& {Umurhan}, O.~M. 2016, \mnras, 456, 3571,
  \dodoi{10.1093/mnras/stv2898}

\bibitem[{{Sch{\"a}fer} \& {Johansen}(2022)}]{SchaferJohansen2022}
{Sch{\"a}fer}, U., \& {Johansen}, A. 2022, \aap, 666, A98,
  \dodoi{10.1051/0004-6361/202243655}

\bibitem[{{Sch{\"a}fer} {et~al.}(2020){Sch{\"a}fer}, {Johansen}, \&
  {Banerjee}}]{SchaferJohansen2020}
{Sch{\"a}fer}, U., {Johansen}, A., \& {Banerjee}, R. 2020, \aap, 635, A190,
  \dodoi{10.1051/0004-6361/201937371}

\bibitem[{{Sch{\"a}fer} {et~al.}(2025){Sch{\"a}fer}, {Johansen}, \&
  {Flock}}]{Schafer2025}
{Sch{\"a}fer}, U., {Johansen}, A., \& {Flock}, M. 2025, arXiv e-prints,
  arXiv:2501.07633.
\newblock \doarXiv{2501.07633}

\bibitem[{{Sekiya}(1998)}]{Sekiya1998}
{Sekiya}, M. 1998, \icarus, 133, 298, \dodoi{10.1006/icar.1998.5933}

\bibitem[{{Sengupta} \& {Umurhan}(2023)}]{SenguptaUmurhan2023}
{Sengupta}, D., \& {Umurhan}, O.~M. 2023, \apj, 942, 74,
  \dodoi{10.3847/1538-4357/ac9411}

\bibitem[{{Simon} {et~al.}(2016){Simon}, {Armitage}, {Li}, \&
  {Youdin}}]{Simon2016}
{Simon}, J.~B., {Armitage}, P.~J., {Li}, R., \& {Youdin}, A.~N. 2016, \apj,
  822, 55, \dodoi{10.3847/0004-637X/822/1/55}

\bibitem[{{Simon} {et~al.}(2013){Simon}, {Bai}, {Stone}, {Armitage}, \&
  {Beckwith}}]{Simon2013}
{Simon}, J.~B., {Bai}, X.-N., {Stone}, J.~M., {Armitage}, P.~J., \& {Beckwith},
  K. 2013, \apj, 764, 66, \dodoi{10.1088/0004-637X/764/1/66}

\bibitem[{{Squire} \& {Hopkins}(2018)}]{SquireHopkins2018}
{Squire}, J., \& {Hopkins}, P.~F. 2018, \mnras, 477, 5011,
  \dodoi{10.1093/mnras/sty854}

\bibitem[{{Squire} \& {Hopkins}(2020)}]{SquireHopkins2020}
---. 2020, \mnras, 498, 1239, \dodoi{10.1093/mnras/staa2311}

\bibitem[{{Stoll} \& {Kley}(2016)}]{StollKley2016}
{Stoll}, M. H.~R., \& {Kley}, W. 2016, \aap, 594, A57,
  \dodoi{10.1051/0004-6361/201527716}

\bibitem[{{Stone} {et~al.}(2020){Stone}, {Tomida}, {White}, \&
  {Felker}}]{Stone2020}
{Stone}, J.~M., {Tomida}, K., {White}, C.~J., \& {Felker}, K.~G. 2020, \apjs,
  249, 4, \dodoi{10.3847/1538-4365/ab929b}

\bibitem[{{Suriano} {et~al.}(2018){Suriano}, {Li}, {Krasnopolsky}, \&
  {Shang}}]{Suriano2018}
{Suriano}, S.~S., {Li}, Z.-Y., {Krasnopolsky}, R., \& {Shang}, H. 2018, \mnras,
  477, 1239, \dodoi{10.1093/mnras/sty717}

\bibitem[{{Tominaga} \& {Tanaka}(2023)}]{Tominaga2023}
{Tominaga}, R.~T., \& {Tanaka}, H. 2023, \apj, 958, 168,
  \dodoi{10.3847/1538-4357/ad002e}

\bibitem[{{Umurhan} {et~al.}(2020){Umurhan}, {Estrada}, \&
  {Cuzzi}}]{Umurhan2020}
{Umurhan}, O.~M., {Estrada}, P.~R., \& {Cuzzi}, J.~N. 2020, \apj, 895, 4,
  \dodoi{10.3847/1538-4357/ab899d}

\bibitem[{{Urpin}(2003)}]{Urpin2003}
{Urpin}, V. 2003, \aap, 404, 397, \dodoi{10.1051/0004-6361:20030513}

\bibitem[{{Urpin} \& {Brandenburg}(1998)}]{UrpinBrandenburg1998}
{Urpin}, V., \& {Brandenburg}, A. 1998, \mnras, 294, 399,
  \dodoi{10.1046/j.1365-8711.1998.01118.x}

\bibitem[{{van der Marel} {et~al.}(2016){van der Marel}, {Cazzoletti},
  {Pinilla}, \& {Garufi}}]{vanderMarel2016}
{van der Marel}, N., {Cazzoletti}, P., {Pinilla}, P., \& {Garufi}, A. 2016,
  \apj, 832, 178, \dodoi{10.3847/0004-637X/832/2/178}

\bibitem[{{van der Marel} {et~al.}(2013){van der Marel}, {van Dishoeck},
  {Bruderer}, {Birnstiel}, {Pinilla}, {Dullemond}, {van Kempen}, {Schmalzl},
  {Brown}, {Herczeg}, {Mathews}, \& {Geers}}]{vanderMarel2013}
{van der Marel}, N., {van Dishoeck}, E.~F., {Bruderer}, S., {et~al.} 2013,
  Science, 340, 1199, \dodoi{10.1126/science.1236770}

\bibitem[{{Weidenschilling}(1977)}]{Weidenschilling1977Aerodynamics}
{Weidenschilling}, S.~J. 1977, \mnras, 180, 57, \dodoi{10.1093/mnras/180.2.57}

\bibitem[{{Weidenschilling}(1980)}]{Weidenschilling1980Planetesimals}
---. 1980, \icarus, 44, 172, \dodoi{10.1016/0019-1035(80)90064-0}

\bibitem[{{Xu} \& {Bai}(2022)}]{XuBai2022}
{Xu}, Z., \& {Bai}, X.-N. 2022, \apj, 924, 3, \dodoi{10.3847/1538-4357/ac31a7}

\bibitem[{{Yang} \& {Zhu}(2020)}]{YangZhu2020}
{Yang}, C.-C., \& {Zhu}, Z. 2020, \mnras, 491, 4702,
  \dodoi{10.1093/mnras/stz3232}

\bibitem[{{Yang} \& {Zhu}(2021)}]{YangZhu2021SI}
---. 2021, \mnras, 508, 5538, \dodoi{10.1093/mnras/stab2959}

\bibitem[{{Yang} {et~al.}(2023){Yang}, {Fern{\'a}ndez-L{\'o}pez}, {Li},
  {Stephens}, {Looney}, {Lin}, \& {Harrison}}]{Yang2023}
{Yang}, H., {Fern{\'a}ndez-L{\'o}pez}, M., {Li}, Z.-Y., {et~al.} 2023, \apjl,
  948, L2, \dodoi{10.3847/2041-8213/acccf8}

\bibitem[{{Youdin} \& {Johansen}(2007)}]{YoudinJohansen2007}
{Youdin}, A., \& {Johansen}, A. 2007, \apj, 662, 613, \dodoi{10.1086/516729}

\bibitem[{{Youdin} \& {Goodman}(2005)}]{YoudinGoodman2005}
{Youdin}, A.~N., \& {Goodman}, J. 2005, \apj, 620, 459, \dodoi{10.1086/426895}

\bibitem[{{Youdin} \& {Lithwick}(2007)}]{YoudinLithwick2007}
{Youdin}, A.~N., \& {Lithwick}, Y. 2007, \icarus, 192, 588,
  \dodoi{10.1016/j.icarus.2007.07.012}

\bibitem[{{Youdin} \& {Shu}(2002)}]{YoudinShu2002}
{Youdin}, A.~N., \& {Shu}, F.~H. 2002, \apj, 580, 494, \dodoi{10.1086/343109}

\bibitem[{{Zhang} {et~al.}(2024){Zhang}, {Zhu}, \& {Jiang}}]{ZhangZhu2024}
{Zhang}, S., {Zhu}, Z., \& {Jiang}, Y.-F. 2024, \apj, 968, 29,
  \dodoi{10.3847/1538-4357/ad4109}

\bibitem[{{Zhu} {et~al.}(2014){Zhu}, {Stone}, {Rafikov}, \&
  {Bai}}]{ZhuStone2014}
{Zhu}, Z., {Stone}, J.~M., {Rafikov}, R.~R., \& {Bai}, X.-n. 2014, \apj, 785,
  122, \dodoi{10.1088/0004-637X/785/2/122}

\bibitem[{{Zhu} \& {Yang}(2021)}]{ZhuYang2021}
{Zhu}, Z., \& {Yang}, C.-C. 2021, \mnras, 501, 467,
  \dodoi{10.1093/mnras/staa3628}

\end{thebibliography}

\end{document}